\documentstyle[proceedings]{crckapb}

\def\be{\begin{equation}}
\def\ee{\end{equation}}
\def\bea{\begin{eqnarray}}
\def\eea{\end{eqnarray}}

\def\sm{M_\odot}
\def\etal{et al. }

\begin{opening}
\title{ RECENT DEVELOPMENTS IN THE SEARCH
 FOR BARYONIC DARK MATTER}

\author{B. J. CARR}
\institute{Astronomy Unit, Queen Mary \& Westfield College, \\
Mile End Road, London E1 4NS, UK}

\end{opening}

\runningtitle{ }

\begin{document}

\begin{abstract}
Cosmological nucleosynthesis calculations imply that many of 
the baryons in the Universe must be dark. We discuss the 
likelihood that some of these dark baryons may reside in galaxies as
Massive Compact Halo Objects (MACHOs), the remnants of a first generation of pregalactic or protogalactic stars. Various candidates
have been proposed for such remnants and we review the
many types of observations
which can be used to detect or exclude them.
Claims to have found positive evidence for some of the candidates
have generally turned out to be spurious or questionable, so the status
of the MACHO scenario remains controversial. However, it would be
premature to reject MACHOs altogether and further
observations are likely to resolve the issue soon.
\end{abstract}

\section{Introduction}
This talk will address four issues: (1) What is the evidence that some of the baryons in the Universe are dark? (2) What are the reasons for believing that some of the dark baryons are in galaxies (i.e. in the form of MACHOs)? (3) How good is the evidence for such objects from microlensing? 
(4) What are the most plausible MACHO candidates?
It may be regarded as a sequel to my contribution at a previous Chalonge school (Carr 1995). Although there have been many interesting developments since then, with the status of the MACHO scenario going through various vacillations, much of the pedagogical discussion of my earlier paper still applies. It should therefore usefully complement  the present paper.

Evidence for dark matter has been claimed in four different 
contexts and it is important to distinguish between them in assessing which can
have a baryonic explanation. There may be {\em local\/} dark matter in the Galactic disc, dark matter in the {\em halos\/} of our own and other galaxies, dark matter associated with {\em clusters\/} of galaxies and finally - if one believes that the total
cosmological density has the critical value - smoothly 
distributed {\em background\/} dark matter. Since dark matter
probably takes as many different forms as visible matter, it would be simplistic
to assume that all these dark matter problems have a single solution. 
The local
dark matter is almost certainly unrelated to the other ones and, while the halo and
cluster dark matter would be naturally connected in many evolutionary scenarios, there is a growing tendency to regard
the unclustered background dark matter as different from the clustered component. 
As discussed by Schmidt (2001) at this meeting, this is because the latest supernovae measurements indicate that the
cosmological expansion is accelerating, which means that the total density must be
dominated by some form of ``exotic'' energy with negative pressure (possibly a cosmological constant) and this would not be expected to cluster.

The combination of the supernovae and microwave background observations suggest
that the density parameters associated with the exotic component and the
ordinary (positive pressure) matter component are $\Omega_X\approx 0.7$ and $\Omega_M=0.3$ respectively (Richards 2000). This is compatible with the total density parameter being $1$, as expected in the inflationary scenario.
The matter density may itself be broken down into different components (Turner 1999). Large-scale structure observations suggest that there
must be {\em cold\/} dark matter (eg. WIMPs) with a density parameter 
$\Omega_C=0.3\pm0.1$, the latest determination of the neutrino mass by the Super-Kamiokande experiment requires that there is {\em hot\/} dark matter with a density parameter $\Omega_H \approx 0.01$, 
and we will see that Big Bang nucleosynthesis calculations require that the baryonic matter (be it visible or dark) has a density parameter
$\Omega_B=0.05\pm0.005$. (All these $\Omega$ values assume a Hubble parameter $H_o =65$~km~s$^{-1}$~Mpc$^{-1}$.) It is remarkable, not only that each of these components seems to be needed, but also that their densities are all 
within one or two orders of magnitude of each other:
\be
1>\Omega_X \sim \Omega_C \sim \Omega_H \sim \Omega_B >0.01.
\ee
Why this should be remains a mystery.
Although we will be focussing henceforth on baryonic dark matter,  it is important to place the considerations that follow in this broader context.

\section{Evidence for Baryonic Dark Matter}
The main argument for both baryonic and
non-baryonic dark matter comes from Big Bang nucleosynthesis calculations. 
This is because the success of the standard picture in explaining the
primordial light element abundances only applies if the baryon 
density parameter lies in the range (Copi \etal 1995)
   \be
      0.007h^{-2} < \Omega_B < 0.022h^{-2}
   \ee
where $h$ is the Hubble parameter in units of
$100$~km~s$^{-1}$~Mpc$^{-1}$. For comparison, the latest measurements of the primordial deuterium abundance imply a much tighter constraint (Burles et al. 1999):
  \be
      0.018h^{-2} < \Omega_B < 0.020h^{-2}.
   \ee
In any case,
the upper limit implies that $\Omega_B$ is well below 1, which suggests that no
baryonic candidate could provide the matter density required by large-scale
structure observations. This conclusion also applies if one invokes
inhomogeneous nucleosynthesis since one requires $\Omega_B<0.09h^{-2}$ even
in this case (Mathews \etal 1993). On the other hand, the value 
of $\Omega_B$
allowed by eqns (2) and (3) almost certainly exceeds the density of 
visible
baryons $\Omega_V$. A careful inventory by Persic \& Salucci 
(1992)
shows that the density in galaxies and cluster gas is 
\be
\Omega_V \approx (2.2 + 0.6h^{-1.5}) \times 10^{-3} \approx 0.003
\ee
where the last approximation applies
for reasonable values of $h$. This is well below the lower limits allowed
by eqns (2) and (3), so it seems that one needs {\em both\/}
non-baryonic and baryonic dark matter.

Can the claim that some of the nucleosynthetic baryons must be dark be circumvented by decreasing $\Omega_B$ or increasing
$\Omega_V$ in some way? A few years ago the anomalously high deuterium abundance measured in some intergalactic Lyman-$\alpha$ clouds suggested that the
nucleosynthesis value for $\Omega_B$ could be lower than the standard one (Carswell et al. 1994, Songaila et al. 1994, Rugers \& Hogan 1996, Webb at al. 1997). However, the evidence for this was always disputed (Tytler et al. 1996) and recent studies of the
quasars Q1937 and Q1009 suggest that the deuterium abundance is $3.3 \times 10^{-5}$ (Burles \& Tytler 1998). This corresponds to $\Omega_Bh^2=0.019$, which is
in the middle of the range given by eqn (3). In fact, recent measurements of the 2nd anisotropy peak in the microwave background suggest $\Omega_Bh^2=0.03$, which is even larger (de Bernardis et al. 2000, Richards 2001).

The possibility that $\Omega_V$ could be larger than indicated by eqn (4) is much
harder to refute. Certainly one can now make a more precise estimate for some of the components of $\Omega_V$ than Persic \& Salucci. A recent review by 
Hogan (1999) replaces the factors in brackets in eqn (4) by 2.6 for spheroid stars, 0.86 for disc stars, 0.33 for neutral atomic gas, 0.30 for
molecular gas and 2.6 for the ionized gas in clusters. This assumes $H_o =70$~km~s$^{-1}$~Mpc$^{-1}$. However, this may still not account for all components.  For example, there may be some baryons in low surface brightness galaxies (De Blok \& McGaugh 1997) or dwarf galaxies (Loveday et al. 1997) or in a hot intergalactic medium. The last possibility is emphasized by Cen \& Ostriker (1999), who
find that - in the context of the CDM scenario - half the mass of baryons should be in warm ($10^5-10^7$K) intergalactic gas at the present epoch. Phillips et al. (2000)
claim that there may be some evidence for this from cosmological X-ray background observations.

Another possibility is that the missing baryons
could be in gas in groups of clusters. This has been emphasized by Fukugita et al. (1998), who argue 
that plasma in groups (which would be too cool to have appreciable X-ray emission)
could provide the cosmological nucleosynthesis shortfall. Indeed the review by Hogan (1999)
suggests that the ionized gas in groups could have a density parameter as high as $0.014$. However, it must be stressed that this estimate is based on an extrapolation of observations of rich clusters and there is no direct evidence for this.

\section{Is there Baryonic Dark Matter in Galaxies?}
Which of the dark matter problems identified in Section 1 could be baryonic? Certainly the dark matter in
galactic discs could be - indeed this is the only dark matter problem which is definitely
baryonic. Even if all discs have the 60\% dark component
envisaged for the Galaxy by Bahcall \etal (1992), this only 
corresponds to $\Omega_d \approx
0.001$, well below the nucleosynthesis bound. However, the Bahcall et al. claim is strongly disputed by Kuijken \& Gilmore (1989) and Flynn \& Fuchs (1995). Indeed recent Hipparcos observations suggest that the dark disc fraction is below 10\% (Cr\'ez\'e \etal 1998).

The more interesting question is whether the 
halo dark matter could be baryonic. If the Milky Way is typical, the density associated with halos would 
be $\Omega_h \approx 0.03h^{-1}(R_h/100\mbox{kpc})$, where $R_h$ is the (uncertain) halo radius, so the upper limit in eqn (2) 
implies that {\em all\/} the dark matter in halos could be baryonic 
providing
$R_h<70h^{-1}$kpc. This is marginally possible for our galaxy (Fich \& 
Tremaine 1991), in which case the dark baryons could be contained in the remnants of a first generation of ``Population III" stars (Carr 1994). This corresponds to the ``Massive Compact Halo Object" or ``MACHO" scenario and has attracted considerable interest as a result of the LMC microlensing observations. Even if halos are larger than $70h^{-1}$kpc, and studies of the kinematics of other spirals by Zaritsky et al. (1993) suggest that they could be as large as 
$200h^{-1}$kpc, a considerable fraction of their mass could still be in stellar remnants.

Possible evidence for dark baryons in galaxies comes from
studying the density profiles and rotation curves for dwarf galaxies. It is well known that the presence of
dark matter in dwarf galaxies requires that it cannot be entirely hot. However,
if the dark matter consisted entirely of WIMPs, one would expect it to have the standard 
Navarro-Frenk-White density profile (Navarro et al. 1997) and Burkert  \& Silk (1999) claim that the profile for
DDO 154 is very different from this. They argue that the measurements indicate the presence of a centrally condensed baryonic dark component, having about 25\% of the total dark matter density. Another possible signature of a baryonic halo would be {\it flattening} of the halo since more dissipation would 
be expected in this case. For our galaxy observations are best fit by an axis ratio $q\approx 0.6$, which
constrains the fraction of the halo in MACHOs but does not
necesssarily require them (Samurovic et al. 1999).
 
On theoretial grounds one would {\it expect} the halo dark matter to be a mixture of WIMPs and MACHOs. For since the cluster dark matter must be predominantly cold, one would expect at least some of it to clump into galactic
halos. The relative densities of the two components would depend sensitively on
the formation epoch of the Population III stars. If they formed pregalactically, one would expect
the halo ratio to reflect the cosmological background ratio (viz. $\Omega_B/\Omega_C \approx 0.1$). However, if they formed protogalactically, the ratio could be larger
since the baryons could have dissipated and become  more centrally concentrated 
than the WIMPs.
 
In order to distinguish between the pregalactic and protogalactic scenarios, it is important to gain independence evidence
about the formation epoch of the putative MACHOs. At moderate redshifts one can obtain a lower limit to the baryon density by studying Lyman-$\alpha$ clouds. The simulations of Weinberg et al. (1997) suggest that the
density parameter of the clouds must be at least $0.017h^{-2}$ at a redshift of 2 and this is already
close to the upper limit given by eqn (3). By today some of these baryons might have been transformed into a hot intergalactic medium or stars but this suggests that there was little room for any dark baryons before $z=2$.

\section{Lensing Evidence}
Gravitational lensing effects could permit the detection of compact
objects over the 
entire mass range $10^{-7}\sm$ to $10^{12}\sm$. The associated constraints 
on the density parameter of compact objects as a function of mass 
are brought together in Figure 1, shaded regions being excluded. This is an 
updated version of the figure presented at the last Chalonge meeting (Carr 1995)
and a more detailed discussion of the limits is provided there. 
Here I will
focus (more positively) on the evidence rather than the constraints but it must be stressed at the outset that this evidence
should not be regarded as definitive. The LMC microlensing events,
for example, are sometimes interpreted as evidence for MACHOs but they have also been attributed to LMC self-lensing (Sahu 1994), a warped or flared Galactic disc (Evans et al. 1998),
an old superthick disc (Gates \& Gyuk 1999), intervening debris
(Zhao 1998) or the LMC's own halo
(Kerins \& Evans 1999). Nevertheless, the following discussion shows that this is not the only context in which microlensing evidence for compact objects arises and it is interesting that the other cases indicate a similar lens mass.
\vspace{10pt}

\begin{figure}\label{F1}
\vskip 3.5in
\caption{Lensing constraints on the density parameter for compact objects}
\end{figure}

{\em Microlensing of Stars in LMC}. 
Attempts to detect microlensing by objects in our own halo by 
looking for
intensity variations in stars in the Magellanic Clouds and the
Galactic Bulge have now been underway for a decade (Paczynski 1996). This method
is sensitive to lens masses in the range $10^{-7}-10^2\sm$ but the probability of an
individual star being lensed is only $\tau \sim 10^{-6}$, so one 
has to
look at many stars for a long time (Paczynski 1986). The duration and likely 
event rate are
$P\sim 0.2(M/\sm)^{1/2}y$ and $\Gamma \sim N \tau P^{-1} \sim (M/\sm)^{-1/2}y^{-1}$
where $N\sim 10^6$ is the number of stars.
The MACHO 
group currently has 13-17 LMC events and the durations span the range 
$34-230$ days (Alcock et al. 
2000a). For a standard halo model, the data suggest an average lens mass of around $0.5~\sm$ and a halo fraction of 0.2, with the 95\% confidence ranges being $0.15-0.9~\sm$ and $0.08-0.5$. The mass is comparable with earlier estimates (Alcock et al. 1997) but the
fraction is somewhat smaller. This
might appear to indicate that the MACHOs are white
dwarfs but, as discussed in Section 5, this may be excluded on astrophysical grounds, so this presents a dilemma for MACHO enthusiasts. 
One possible resolution is to invoke a less conventional candidate; for
example, primordial black holes forming at the
quark-hadron phase transition might have the required mass 
(Jedamzik 1997). Perhaps the
most important result of the LMC searches is that they {\em eliminate} many candidates. Indeed the combination of the MACHO and EROS results already excludes objects in the mass range $5\times10^{-7}-0.002~\sm$ from having more than 0.2 of the halo density (Lasserre et al. 2000). One can also exclude objects in the larger mass range $0.3-30~\sm$ from having the full halo density (Alcock et al. 2000b).
\vspace{10pt}

{\em Microlensing of Stars in M31}.
The LMC studies are complemented by searches for microlensing of stars in M31.
In this case, the sources are too distant to resolve individually (i.e. there are many
stars per pixel), so a lensing event
is observed only if the amplification is large enough for the source to stand out 
above the background, but observations of
the LMC already demonstrate the efficacy of the method (Melchior et al. 1999). For sources in M31 the halo objects may reside in either our own galaxy or M31 but  the crucial point
is that one expects an asymmetry between the far and near side of the disc. Two
groups have been involved in this work: the AGAPE collaboration (Ansari et al. 1997),
who use the ``pixel" method, and the VATT-Columbia collaboration, who use
``differential image photometry" (Crotts \& Tomaney 1997). The AGAPE team have been monitoring seven fields
in M31 in red and blue and have already detected one good lensing candidate with a likely mass of  $0.6~\sm$
(Ansari et al. 1999). The theoretical importance of detecting microlensing in M31 is discussed in detail by Kerins et al. (2000).
\vspace{10pt}

{\em Microlensing in Macrolensed Sources}. If a galaxy is suitably positioned to 
image-double a quasar, then there is also a high probability that 
an individual halo object will traverse the line of sight of one of the
images. This will give intensity fluctuations in one
but not both images (Gott 1981), the timescale of the fluctuations being
$\sim 40(M/\sm)^{1/2}$y. There is already evidence of this effect for the 
quasar
2237+0305 (Irwin \etal 1989), the observed timescale
possibly indicating a mass below $0.1~\sm$ (Webster \etal 1991). However, because
the optical depth is high, the mass estimate is rather uncertain and
a more recent analysis suggests that it could be in the higher range
$0.1-10~\sm$ (Lewis et al. 1998), in which case the lens could be an
ordinary star. The absence of this effect in the quasar 0957+561 has also been used to exclude MACHOs with mass in the range $10^{-7} - 10^{-3}\sm$ from making up all of the halo of the intervening galaxy (Schmidt \& Wambsganss 1998). Another application of this method
is to seek microlensing in a compact radio source which is macrolensed by a galaxy. Koopmans \& de Bruyn (2000) claim to have detected this effect for the 
CLASS gravitational lens B1600+434. The inferred lens mass is around $0.5~\sm$, comparable to the mass implied by the LMC data.
\vspace{10pt}

{\em Lensing of Quasars by Dark Objects in Clusters}.
If halos are made of MACHOs,  one would also expect some of these to be spread throughout a cluster of galaxies. This is because individual galaxies
should be stripped of some of their outer halo as a result of collisions and tidal interactions. This method is sensitive to MACHOs in the
mass range $10^{-6} - 10^{-3}~\sm$. Tadros et al. (1998) have therefore looked for the microlensing of quasars by MACHOs in the Virgo
cluster: four months of observations of 600 quasars with the UK Schmidt telescope have yielded no candidates and this already implies that less than half
the mass of Virgo is in $10^{-5}~\sm$ objects. A more extensive follow-up
campaign is currently underway.
\vspace{10pt}

{\em Microlensing of Quasars}. More dramatic 
but rather controversial evidence for the microlensing of quasars 
comes from Hawkins (1993, 1996, 1999), who has been monitoring 300 quasars in 
the redshift range $1-3$ for nearly 20 years using a wide-field 
Schmidt camera. He finds quasi-sinusoidal variations with an amplitude of 
0.5 magnitudes on a timescale 5~y and attributes this to lenses with 
mass $\sim 10^{-3}\sm$. The crucial point is that the timescale 
decreases with
increasing redshift, which is the opposite to what one would 
expect for intrinsic variations. The timescale also increases with the luminosity of the quasar and he explains this by noting that the
variability timescale should scale with the size of the accretion 
disc (which should itself correlate with luminosity). A rather worrying
feature of Hawkins' claim is that he requires the density of the
lenses to be close to critical (in order that the sources are
transited continuously). This may exclude baryonic lenses, so he invokes primordial black 
holes which form at the quark-hadron phase transition 
(Crawford \& Schramm 1982). 
Of course, Hawkins' lenses are much smaller than the ones required by
the previously mentioned microlensing observations and it would perhaps be strange 
to have two distinct populations. As discussed in Section 5, Walker (1999) has proposed that Hawkins' lenses might also be jupiter-mass gas clouds.
\vspace{10pt}

{\em Microlensing of Supernovae by Halos}.
In principle, galactic halos could produce luminosity variations
in high redshift supernovae, many of which are now routinely detected as
part of the Supernova Project. As pointed out by Metcalf \& Silk (1999), a
particularly interesting aspect of this effect is that it could discriminate between
MACHO and WIMP halos. This is because the distribution of amplifications
would be different for what they term ``macroscopic" and ``microscopic" dark matter,
although the shape of the distribution is also sensitive to the cosmological and halo
model.

\section{Assessing the MACHO Candidates}
Although it is not definite that MACHOs exist, one can already place 
interesting constraints on the possible candidates. These are indicated in Figure 2, which
shows which candidates are excluded by various types of observational signature.
Some of these constraints are discussed in more detail in
Carr (1995). A cross indicates that exclusion is definite, while a question mark indicates that it is tentative. Candidates associated with one or more crosses should clearly be rejected but
those with question marks alone may still be viable.
Although no candidate is entirely free of crosses or questions marks, the title of a
recent paper by Freese et al. (2000), ``Death of Baryonic Dark Matter", may be overly pessimistic. We will now discuss each of the candidates in turn.

\begin{figure}\label{F2}
\vskip 3.0in
\caption{Constraints on MACHO candidates}
\end{figure}

\subsection{Brown Dwarfs}
Objects in the range $0.001-0.08~\sm$ would never burn hydrogen and are termed ``brown dwarfs" (BDs). They represent a balance between gravity and degeneracy pressure. Objects below $0.001~\sm$, being held together by intermolecular rather than gravitational forces, have atomic density and are sometimes termed ``snowballs" (SBs). However, such objects would have evaporated within the age of the Universe if they were smaller than $10^{-8}\sm$ (De Rujula et al. 1992) and there are various dynamical constraints for snowballs larger than this (Hills 1986). 

It has been argued that objects below the hydrogen-burning 
limit  may form efficiently in pregalactic or protogalactic cooling flows (Ashman \& Carr 1990, Thomas \& Fabian 1990) but the direct evidence for such objects
remains weak. While some BDs have been found as
companions to ordinary stars, these can only have a tiny cosmological density and 
it is much harder to find isolated field BDs. The best argument therefore comes from 
extrapolating the initial mass function (IMF) of hydrogen-burning stars to lower masses than can be observed directly. The IMF for Population I stars ($dN/dm \sim m^{-\alpha}$ with $\alpha<1.8$) suggests that only 1\% of the disc could be in BDs (Kroupa et al. 1993). However, one might wonder whether $\alpha$ could be larger, increasing the BD fraction, for zero-metallicity stars. Although there are theoretical reasons for entertaining this possibility, earlier observational claims that low metallicity  objects have a steeper IMF than usual are now discredited. Indeed observations of 
Galactic and LMC globular clusters (Elson et al. 1999) and dwarf spheroidal field stars
(Feltzing et al. 1999) suggest that the IMF is {\it universal} with $\alpha <1.5$ at low masses (Gilmore 1999). This implies that the BD fraction is much less than 1\% by mass. However, it should be stressed that nobody has yet measured the IMF in the sites which are most likely to be associated with Population III stars.

We have seen that the LMC microlensing results would seem to exclude a large fraction of BDs in our own halo. Although Honma \& Kan-ya (1998) have proposed 100\% BD models, these would require falling rotation curves and most theorists would regard these as rather implausible.  Another exotic possibility, suggested by Hansen (1999), is ``Beige Dwarfs" in the
mass range  $0.1-0.3~\sm$. Such objects are larger than the traditional BD upper limit but they are supposed to
form by sufficiently slow accretion that they never ignite their nuclear fuel.

\subsection{Red Dwarfs}
Discrete source counts for our own Galaxy suggest that the fraction of the halo mass in low mass hydrogen burning stars - red dwarf (RDs) - must be less than 1\% (Bahcall et al. 1995, Gould et al. 1998, Freese et al. 2000). These limits might be weakened if the stars were clustered
(Kerins 1997) but not by much. For other galaxies, the best constraint on the red dwarf fraction comes from upper limits on the halo red light and such studies go back 
several decades (Boughn \& Saulson 1983). 

The discovery of a red light around NGC 5907 by Sackett et al. (1994), apparently emanating from low mass stars with a density profile like that of the halo, was therefore a particularly interesting development. This detection was confirmed in V and I by Lequeux et al. (1996) and in J and K by James \& Casali (1996). However, the suggestion that
the stars might be of primordial origin (with low metallicity) was contradicted by the results of Rudy et al. (1997), who found that the color was indicative of low mass stars with solar metallicity.
Furthermore, it must be stressed that the red light light has only been observed within a few kpc and no NIR emission is detected at 10-30 kpc (Yost et al. 1999). Both these points go against the suggestion that the red light is associated with MACHOs.

Recently it has been suggested that the red light seen
in NGC 5907 is more likely to derive from a disrupted dwarf galaxy, the stars of which
would naturally follow the dark matter profile (Lequeux et al. 1998), or to be a ring left over from a disrupted dwarf spheroidal galaxy (Zheng et al. 1999). However, in this case one would expect of order a hundred bright giants for a standard IMF, whereas NICMOS observations find only one (Zepf et al. 2000). There is clearly still a puzzle here. In any case, NGC 5907 does not seem to be typical since ISO observations of four other edge-on bulgeless spiral galaxies give no evidence for red halos (Gilmore \& Unavane 1998).

\subsection{White Dwarfs}
For many years white dwarfs (WDs) have been regarded as rather implausible dark matter candidates. One requires a very contrived IMF, lying between $2~\sm$ and $8~\sm$, 
in order to avoid excessive production of light or metals (Ryu et al. 1990); the fraction
of WD precursors in binaries is expected to produce too many type 1A supernovae (Smecker \& Wyse 1991); and the halo fraction is constrained to be less than 10\% in order to avoid the luminous precursors contradicting the upper limits from galaxy counts (Charlot \& Silk 1995).  The observed WD luminosity function
also placed a severe lower limit on the age of any WDs in our own halo (Tamanaha et al. 1990). 

More recent constraints have strengthened these limits. A study of CNO production suggests that a halo comprised entirely of WDs would overproduce C and N compared to O by factors as
large as 60 (Gibson \& Mould 1997) and a similar limit comes from considering helium and deuterium production (Field et al. 2000). Extragalactic
background light limits now require that the halo WD fraction be less than 6\% (Madau \& Pozzetti 1999) and 
the detection of TEV $\gamma$-rays from the the galaxy Makarian 501 (which 
indirectly constrains the infrared background) requires that the WD density parameter
be less than $0.002h^{-1}$ (Graff et al. 1999).

The ``many nails in the coffin"  of the WD scenario are confounded by the results of the LMC microlensing observations,
which we have seen suggest that the lens mass is most probably in the WD range. Not surprisingly, therefore, theorists have been trying to resuscitate the scenario. At least some of the afore-mentioned limits (though not the nucleosynthetic ones) must be reconsidered in view of recent claims by
Hansen (1998) that old WDs with pure-hydrogen envelopes could be much bluer and brighter than previously supposed, essentially because the light emerges from deeper in the atmosphere.
This suggestion has been supported by HST observations of Ibata et al. (1999), who claim to have detected five candidates of this kind. The objects are blue and isolated
and show high proper motion. They infer that they are $0.5~\sm$ hydrogen-atmosphere WDs with an age of around $12$ Gyr. Three such objects have now been 
identified spectroscopically (Hodgkin et al. 2000, Ibata et al. 2000), so this possibility
must be taken very seriously. Evidence that the blue population is Galactic comes from the observed asymmetry between the North and South HST fields (Mendez \& Minniti 2000); this is expected from dust obscuration. Another interesting signature of the WD scenario would be the generation of background gravitational waves (Hiscock et al. 2000).

\subsection{Black Hole Remnants}
Stars bigger than about $8~\sm$ would leave neutron star (NS) remnants, while
those in some range above about $20~\sm$ would leave black hole remnants. However, neither of these would be plausible candidates for either the disc or halo dark matter because their precursors would have unacceptable nucleosynthetic yields. 
Stars larger than $200~\sm$ are termed ''Very Massive Objects" or VMOs and might 
collapse to black holes without these nucleosynthetic consequences (Carr et al. 1984). However, during their main-sequence phase, such VMOs would be expected to generate of a lot of background light. By today this should have been shifted into the infrared or submillimetre band, as a result of either redshift effects or dust reprocessing, so one would expect a sizeable infrared/submillimetre cosmic background (Bond et al. 1991). Over the last few decades there have been 
several reported detections of such a background but these have usually turned out to be false alarms. COBE does now seem to have detected a
genuine infrared background (Fixsen et al. 1998) but this can probably be attributed to 
ordinary Population I and II stars. In any case, the current constraints on such a background strongly limit the density of any VMOs and, if they provide an appreciable fraction of halos, they would certainly need to form at a very high (pregalactic) redshift.

Stars larger than $10^5~\sm$ - termed ``Supermassive Objects" or SMOs - would collapse directly to black holes without any nucleosynthetic or background light 
production. However, supermassive black holes are would still have noticeable
lensing effects (as indicated in Figure 2) and dynamical effects. The latter are reviewed in detail by Carr \& Sakellariadou (1998) and are summarized in
Figure 3. The shaded regions indicate the constraints on the density parameter of black holes in various locales: in our own disc due to the disruption of open clusters; in our own halo due to the heating of disc stars, the disruption of globular clusters and dynamical friction effects; in clusters of galaxies due to the disruption of galaxies; and in intergalactic space due to the inducement of
peculiar motions. Although it has been claimed that there is positive evidence for some of these effects, such as disc heating (Lacey \& Ostriker 1985), the interpretation of this evidence is is not clear-cut. 
It should be
stressed that many of the limits in Figure 3 would also apply if the black holes were replaced by ``dark clusters"
of smaller objects, a scenario which has been explored by many authors (Carr \& Lacey 1987, Ashman 1990, Kerins \& Carr 1994, De Paolis et al. 1995, Moore \& Silk 1995).

\begin{figure}\label{F3}
\vskip 3.5in
\caption{Dynamical constraints on compact objects in various locales}
\end{figure}

\subsection{Cold Clouds}
The suggestion that the halo dark matter could be in cold clouds was first made by 
Pfenniger et al. (1994). They envisaged the clouds having a mass of around $10^{-3}~\sm$
and being distributed in a disc, which now seems dynamically rather implausible, but
De Paolis et al. (1995) have proposed a similar scenario with a spheroidal halo of clouds.
Walker (1999) argues that such clouds could explain both the ``Extreme Scattering Events" detected by radio observations in our own galaxy and the quasar
microlening events claimed by Hawkins. Indeed Walker \& Wardle (1999) advocate a model in which the halo entirely comprises such clouds, with visible stars being formed as a
result of collisions between them. They claim that this scenario naturally produces various observed features of galaxies. Although clouds of  $10^{-3}~\sm$ could not
explain the LMC microlensing events, Draine (1998) argues that such clouds
might still produce the apparent microlensing events through refraction effects. The interaction of cosmic rays with such clouds would also produce an interesting
$\gamma$-ray signature (De Paolis et al. 1999, Sciama 2000). 

\subsection{Primordial Black Holes}
As discussed in more detail in the accompanying paper at this meeting, black holes may have formed in the early Universe, either from initial inhomogeneities or at some sort of cosmological phase transition (Carr 1994). Those forming at time $t$ after the Big Bang would have a mass of order the horizon mass at that epoch $\sim10^5(t/s)\sm$. Since there could not have been large-amplitude horizon-scale inhomogeneities at the epoch of cosmological nucleosynthesis ($t\sim1$~s), PBHs forming via the first mechanism are unlikely to be larger than $10^5\sm$. On the other hand, since there is no phase transition after the quark-hadron era at $10^{-5}s$, those forming via the second are unlikely to be larger than $1~\sm$. The possibility that PBHs may have formed at the quark-hadron transition (Crawford \& Schramm 1982) has attracted considerable attention recently (Jedamzik 1997) because they would naturally have a mass comparable to that required by the microlensing candidates. 
However, this scenario requires fine-tuning since the fraction of the
Universe going into black holes at this transition must only be about $10^{-9}$. Note that, strictly speaking, PBHs should not be described as ``baryonic" since they form during the radiation-dominated era. 

\section{Conclusions}
It would be premature to either reject or accept the MACHO scenario but recent developments lead to various conclusions.

1) Although cosmological nucleosynthesis calculations suggest that many baryons are dark, one cannot be  sure that the dark baryons are inside galaxies. The more
conservative conclusion would be that they are contained in an intergalactic medium or in gas within groups or clusters of galaxies.

2) Over the years there have been several observational claims of effects which seemed to indicate the existence of MACHOs: in particular, disc-heating by halo SMOs, infrared background radiation from VMOs, halo light from red dwarfs. However, these have usually turned out to be false alarms.

3) Currently the only {\it positive} evidence for MACHOs comes from microlensing observations. The LMC results suggest that white dwarfs are currently the best MACHO candidates but the large number of arguments which have been voiced against white dwarfs in the past cannot be brushed aside too cavalierly. The detection of microlensing in M31 and a compact radio source also gives a mass in the white dwarf range but Hawkins' result requires a much smaller mass.

4) We have seen that there are many important {\it constraints} on MACHO candidates and these are summarized in Figure 1. Until there is a definite detection, therefore, the best strategy is to proceed by {\it eliminating} candidates. However, before pronouncing the MACHO scenario ``dead", it is important to stipulate what is meant by 
this. It is already clear that MACHOs do not provide {\it all} the mass in halos, so the
question is how small does the fraction need to be before the scenario is regarded as
uninteresting. Observations will never be able to exclude there being {\it some} MACHOs.

5) What is clearly missing from current speculation is a good cosmological scenario for the formation of the MACHOs. There is considerable uncertainty as to whether they form pregalactically
(as suggested by background light constraints) or more recently (as suggested by 
observations of Lyman-$\alpha$ clouds). There is also the issue of whether they need
to be distributed in a spherical halo or a superthick disc.

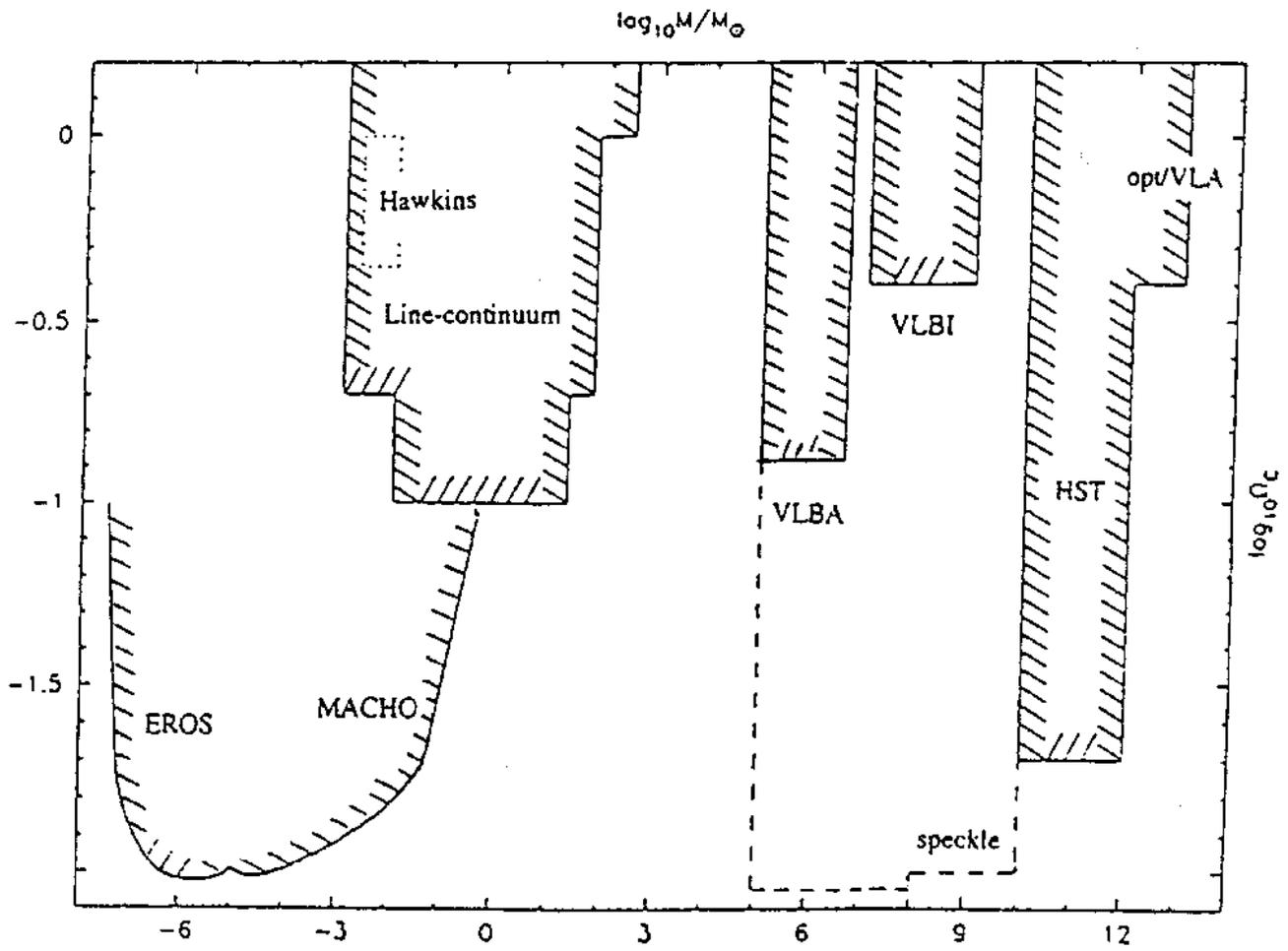
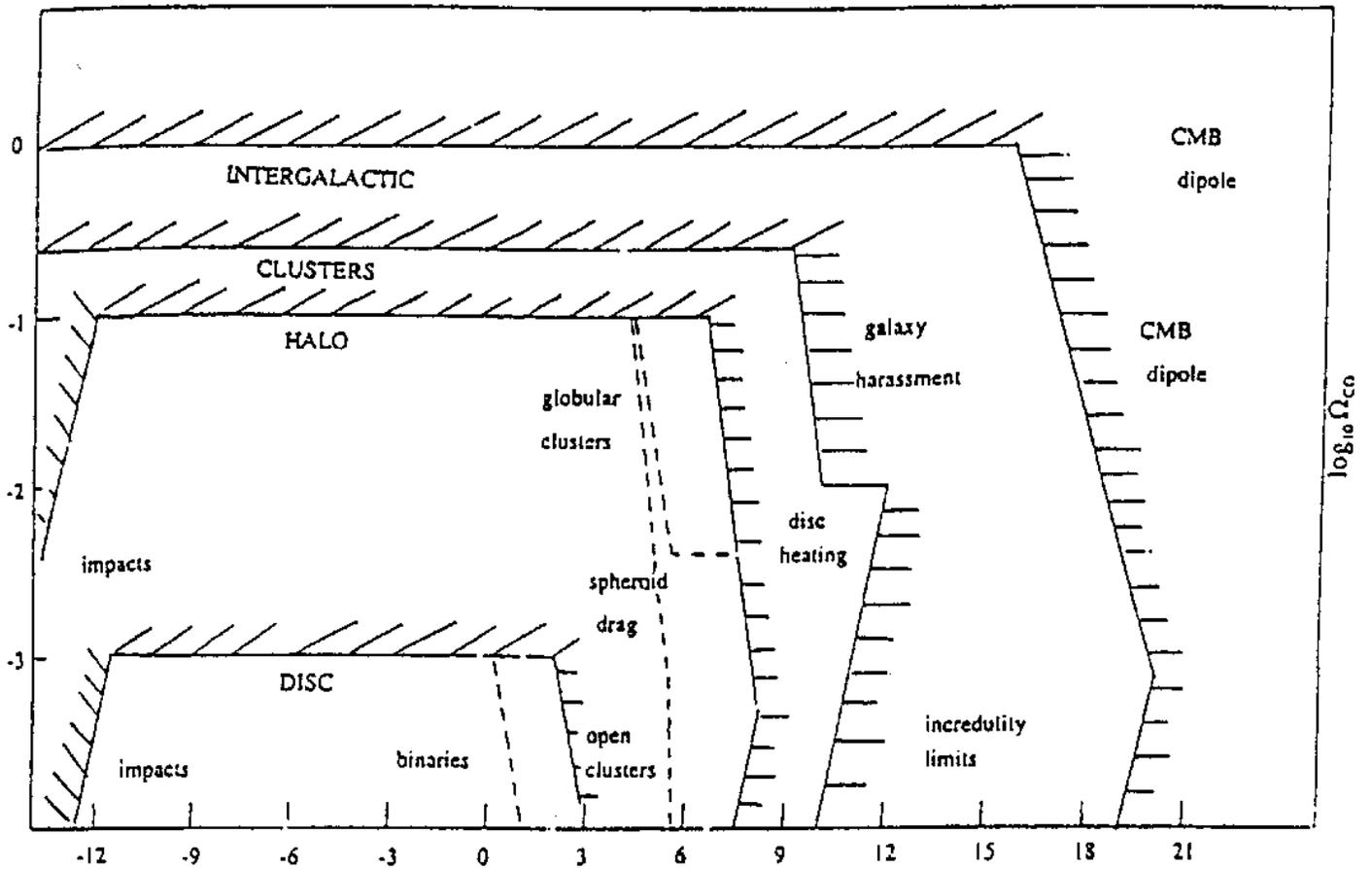
\end{document}